\documentclass[fleqn,twoside,onecolumn,nofootinbib]{revtex4} 

\usepackage[sec,nocpr]{ujp}
\usepackage{graphicx}
\usepackage{dcolumn}
\usepackage{bm}
\usepackage{amssymb,epsfig}
\usepackage{amsfonts}
\usepackage{amsmath}
\usepackage{graphics}

\def\I2{{\large \rm \"{I}}}

\def\i2{\mbox{\footnotesize\rm  \"l\hspace*{-2.45pt}l}}

\def\bi2{\mbox{\footnotesize\rm \bf \"l\hspace*{-2.75pt}l}}

\newfont{\cyrN }{wncyr10 scaled 1150}

\newfont{\cyrS }{wncyr8 scaled 1150}

\newfont{\cyrB }{wncyb10 scaled 1250}

\newfont{\cyrI }{wncyi8 scaled 1000}

\begin{document}

\title{Separate chemical
freeze-outs of strange and non-strange hadrons and  problem of  residual chemical non-equilibrium of strangeness in relativistic heavy ion collisions}

 \author{K.A.~Bugaev}
\affiliation{\bitp}
\address{\bitpaddr}
\email{bugaev@th.physik.uni-frankfurt.de}
\author{D.R. Oliinychenko}
\affiliation{\bitp}
\address{\bitpaddr}
\affiliation{FIAS,Goethe-University, }
\address{Ruth-Moufang Str. 1, 60438 Frankfurt upon Main, Germany}
\email{dimafopf@gmail.com}
 \author{V.V.~Sagun}
\affiliation{\bitp}
\address{\bitpaddr}
\email{v_sagun@ukr.net}
\author{A.I. Ivanytskyi}
\affiliation{\bitp}
\address{\bitpaddr}
\email{a_iv_@ukr.net}
\author{J. Cleymans}
\affiliation{Department of Physics, University of Cape Town}
\address{Rondebosch 7701, South Africa}
\email{jean.cleymans@uct.ac.za}

\author{E.S. Mironchuk}
\affiliation{Moscow Institute of Physics and Technology}
\address{Dolgoprudnyi, 141700 Moscow Region,  Russia}
\email{}

\author{E.G. Nikonov}
\affiliation{Laboratory for Information Technologies, JINR}
\address{141980 Dubna, Russia}
\email{e.nikonov@jinr.ru}

\author{A.V.  Taranenko}
\affiliation{National Research Nuclear University ``MEPhI'' (Moscow Engineering Physics
Institute)}
\address{Kashirskoe Shosse 31, 115409 Moscow, Russia}
\email{}

\author{G.M.  Zinovjev}
\affiliation{\bitp}
\address{\bitpaddr}

 \udk{539.12}  \razd{\seci}

\setcounter{page}{659}%
\maketitle

\begin{abstract}
We present  an elaborate version of the hadron resonance gas model with the combined treatment of separate chemical  freeze-outs for strange and non-strange hadrons and with an additional $\gamma_{s}$ factor which accounts for the remaining  strange particle non-equilibration.  Within suggested approach  the parameters of  two chemical freeze-outs  are connected by the conservation laws of entropy, baryonic charge, third  isospin projection and strangeness. The developed model  enables us to perform a high-quality fit of the hadron multiplicity ratios measured at AGS, SPS and RHIC  with  $\chi^2/dof \simeq 0.93$. A special attention is paid to  a successful  description of  the  Strangeness Horn.  The  well-known  problem
of  selective suppression of  $\bar \Lambda $ and $\bar \Xi$ hyperons   is  also discussed.
The main result is that for all collision energies the $\gamma_{s}$ factor is about 1 within the error bars, except for
the center of mass collision energy 7.6 GeV at which we find about 20\% enhancement of strangeness.
 Also we confirm  an existence of strong jumps  in   pressure,  temperature  and    effective number of degrees of freedom at the stage of  strange particle chemical freeze-out, when the center of mass collision energy changes from 4.3 to 4.9 GeV. We argue that these irregularities
may signal about the quark-hadron phase transition.
\end{abstract}


\section{Introduction}

Relativistic nucleus-nucleus (A+A) collisions provide us with experimental information about the   phase diagram of quantum chromodynamics (QCD) and the strongly interacting matter properties.  The last stage of such collisions is traditionally analyzed within the statistical approach which gives us an excellent opportunity to reveal  the parameters of chemical freeze-out. This approach is based on the assumption of the  thermal equilibrium existence during  the last stage of A+A reaction. Such an equilibrium can be reached due to intensive particle scattering. The stage of the system evolution when the inelastic reactions between hadrons stop is referred to as a  chemical freeze-out (FO). Particle yields are determined by the parameters of FO, namely by chemical potentials and temperature. This general picture is a basis of the Hadron Resonance Gas Model (HRGM) \cite{KABAndronic:05} which is the most successful one in describing the  hadronic  yields measured in heavy-ion experiments for energies from AGS to LHC. Despite a significant success of the HRGM,  in the  experimental data analysis there are a few  unresolved problems. In general they are related to the description of hadron yields which contain (anti)strange quarks. Thus,  the energy dependence of $K^{+}/\pi^{+}$ and $\Lambda/\pi^{-}$ ratios remained  out of high quality description for almost a decade. Excess of strange hadrons yields within the HRGM  led physical community to ponder over strangeness suppression in heavy ion collisions.
The first receipt to resolve this problem was to introduce the strangeness suppression  factor $\gamma_{s}$ which should be fitted in order to describe the experimental data \cite{Rafelsky:gamma}.  However, such an approach is not supported by any underlying physical model and the physical meaning of $\gamma_{s}$ remains unclear \cite{KABAndronic:05,KABAndronic:09,MultiComp:13,KABugaev:Horn2013,KABugaev:tawfik2013,Bugaev13,BugaevIndia}. In addition the strangeness suppression approach in its original form does not contain a hard-core repulsion between hadrons, while the latter is an important feature of the HRGM. A  significant role
of the hard-core repulsion  was demonstrated once more  in Ref. \cite{KABugaev:Horn2013} where the global fit of hadron yield ratios was essentially improved (to $\chi^2/dof \simeq$ 1.16)  compared to all previous analyses.

The most advanced way to account for the hard-core repulsion between hadrons is to consider a hadron gas as a multi-component mixture of particles with different hard-core radii  \cite{KABugaev:Horn2013,KABugaev:tawfik2013,Bugaev13,Bugaev14,MultiComp:08,MultiComp:13}. Within  this approach   all baryons and mesons except for the kaons and the pions are endowed by the common hard-core radii $R_{b}$ and $R_{m}$,  respectively. At the same time the kaon and the  pion radii $R_{K}$ and $R_{\pi}$ are fitted independently in order to provide the best description of $K^{+}/\pi^{+}$ ratio \cite{KABugaev:Horn2013}. This  is an important
finding since  the non-monotonic energy dependence of  $K^{+}/\pi^{+}$ ratio may indicate some  qualitative changes of the system properties and may  serve as a signal of the deconfinement onset. This is a reason why such a  ratio known as  the Strangeness Horn  is of  a special  interest. Note, that the multi-component approach substantially increased the  Strangeness Horn description quality, without spoiling the other ratios including
$\Lambda/\pi^{-}$ one. However, even this advanced approach  does not reproduce the topmost  point of the Strangeness Horn indicating that the  data description is still not ideal. In order to resolve this problem in Ref.  \cite{Bugaev13} the $\gamma_{s}$ factor  was considered as a free parameter within the HRGM with
multi-component repulsion.
Although  the $\gamma_{s}$ data  fit    sizably improves the quality  of Strangeness Horn description,   it does not seem to be useful for  the description of other hadron multiplicities \cite{Bugaev13,Bugaev14}.
 Furthermore, in contrast to the  claims  established on the low-quality fit \cite{Becattini:gammaHIC},  at low energies it was found \cite{Bugaev13} that within the error bars in heavy ion collisions there is an enhancement of strangeness, i.e. $\gamma_{s} >1$,  and not a suppression.   The strangeness enhancement was
 confirmed very recently  \cite{Sagun:2014} by the high quality fit of the available data within the multicomponent HRGM in which the hard-core radius of  $\Lambda$ (anti)hyperon was considered as a global fitting parameter in addition to the set of  hard-core radii used in \cite{KABugaev:Horn2013,KABugaev:tawfik2013,Bugaev13,MultiComp:13,Bugaev14}.

However,  the effect of apparent strangeness non-equilibration can be more successfully explained by the hypothesis of separate chemical FO for all strange hadrons. Since all the hadrons made of $u$ and $d$ quarks are under thermal equilibration whereas the  hadrons containing  $s$ quark are not,   then it is reasonable to assume two different FOs for these two kinds of particles. Following this  conclusion in Ref. \cite{Bugaev13, BugaevIndia} a separate strangeness FO (SFO) was introduced. Note, that according to  \cite{Bugaev13}  both FO and SFO parameters are  connected by the conservation laws of entropy, baryonic charge and isospin projection, while
the net strangeness is explicitly set  to zero at FO and at SFO.
These conservation laws are  crucial elements of the concept of separate SFO developed in \cite{Bugaev13}
which allows  one to essentially reduce the number of  independent fitting parameters.
Another principal element that differs the HRGM of  \cite{Bugaev13,Bugaev14} from the ideal gas  treatment  used in
\cite{BugaevIndia,BugaevIndia2}  is the presence of multi-component hard-core repulsion.

Using the HRGM of    \cite{Bugaev13} it was possible to successfully describe
all hadron multiplicities measured in A+A collisions at AGS, SPS and RHIC energies   with $\chi^2/dof \simeq$ 1.06. The concept of separate SFO led to a systematic improvement of all experimental data description. However, the topmost  point of the Strangeness Horn again was not fitted even within the experimental error, although   the general  description of $K^+/\pi^+$ ratio energy dependence was rather good except  for the upper point.

Since an introduction of the $\gamma_s$ factor demonstrated a remarkable description of all  points of the Strangeness Horn, whereas the separate SFO led to a systematic improvement of all hadron  yields description,  we decided to combine these elements in order to describe an experimental data with the highest possible quality.
{In this way we would like to examine the problem whether the concept of separate SFO is able to completely explain  a possible non-equilibrium of strange charge  and whether
on top of the SFO there exist a necessity to employ the $\gamma_{s}$ factor in statistical approach.}
Note that  from the academic point of view
the problem of  residual strangeness non-equilibration, i.e.
 the question whether the strange charge is or is not in full chemical equilibrium,  is of principal importance.
This ambitious task is the main goal  of the present paper.
Evidently, the best tool for  such a  purpose is  the most successful  version of the HRGM, i.e. the HRGM  with the multi-component hadronic repulsion and a separate SFO.
As it will be shown below,  such an  approach allows us  to describe  111 independent  hadron yield ratios measured for 14 values of the center of mass collision energy $\sqrt{s_{NN}}$ in the interval from 2.7 GeV to  200 GeV with very high  quality.

The paper is organized as follows. The basic features of the developed  model are outlined in  Section 2, while the fitting procedure of the present model is outlined in  Section 3. The main obtained results are  compared with the other models in Section 4. In Section 5 we  discuss in detail  the new fit of  hadronic multiplicity ratios   with two chemical freeze-outs and  $\gamma_{s}$ factor, while Section 6 contains our conclusions.

\section{Model description}

In what follows we treat a hadronic system as a multi-component Boltzmann gas of hard spheres. The effects of quantum statistics are negligible for typical temperatures of the hadronic gas whereas the hard-core repulsion between the particles significantly affects a corresponding equation of state \cite{KABugaev:Horn2013,MultiComp:08}. The present model is dealing with the Grand Canonical treatment. Hence a thermodynamic  state of   system under consideration is fixed by the volume $V$, the temperature $T$, the baryonic
chemical potential $\mu_B$, the strange chemical potential $\mu_S$ and the chemical potential of the  isospin third component $\mu_{I3}$. These parameters control the pressure $p$ of the system. In addition they define the densities $\rho_i^K$ of corresponding charges $Q_i^K$ ($K\in\{B,S,I3\}$).  Introducing the symmetric matrix of the second virial coefficients  $b_{ij} = \frac{2\pi}{3}(R_i+R_j)^3$,
we can obtain the parametric equation of state of the present model in a compact form
\begin{eqnarray}\label{EqI}
p =  \sum_{i=1}^N p_i \,, ~~\rho^K_i = \frac{ Q_i^K{p_i}}{\textstyle  T + \frac{ \sum_{jl} p_j b_{jl} p_l }{p}  }    \,,
\end{eqnarray}
using the partial pressure $p_i$ of $i$-th sort of particles.
The equation of state is written in terms of the solutions $p_i$ of the following system
\begin{eqnarray}\label{EqII}
&&\hspace*{-4mm}p_i =T \phi_i (T)\,   \exp\Biggl[
\frac{\mu_i-2\sum_j p_jb_{ji}+\sum_{jl}p_j b_{jl}p_l/p}{T}
\Biggr]
 \,, \quad \quad \\
&&\hspace*{-4mm}\phi_i (T)  = \frac{g_i}{(2\pi)^3}\int \exp\left(-\frac{\sqrt{k^2+m_i^2}}{T} \right)d^3k  \,.
%
\end{eqnarray}
Each $i^{th}$ sort is characterized by  its  full chemical potential $\mu_i=Q_i^B\mu_i^B+Q_i^S\mu_i^S+Q_i^{I3}\mu_i^{I3}$, mass $m_i$ and degeneracy $g_i$. The function $\phi_i(T)$ denotes the  corresponding particle thermal density in case of ideal gas.
The obtained model parameters for two freeze-outs and their dependence on  the collision energy are discussed in the next section. They are obtained  for the  following values of hard-core radii which were determined  earlier in \cite{KABugaev:Horn2013,Bugaev13}:  $R_{\pi} =0.1$ fm for pions, $R_{K}=0.38$ fm for kaons, $R_{m} = 0.4$ for all other mesons  and $R_{b}=0.2$ fm for all baryons.

In order to  account for the possible  strangeness non-equilibration we introduce the  $\gamma_s$ factor in a conventional way by replacing $\phi_i$ in Eq. (\ref{EqII}) as
\begin{eqnarray} \label{eq:gamma_s}
\phi_i(T) \to \phi_i(T) \gamma_s^{s_i} \,,
\end{eqnarray}
where $s_i$ is a number of strange valence quarks plus number of strange  valence anti-quarks.

The principal  difference of the present  model from the  traditional approaches is that we employ an independent chemical FO of strange particles. Let us consider
this in some detail. The independent freeze-out of strangeness means that inelastic reactions (except for the decays) with  hadrons made of s quarks are switched off at the temperature $T_{\rm SFO}$, the baryonic chemical potential $\mu_{B_{\rm SFO}}$, the strange chemical potential
$\mu_{S_{\rm SFO}}$, the isospin third projection chemical potential $\mu_{I3_{\rm SFO}}$ and the three-dimensional emission volume $V_{\rm SFO}$. In general case these parameters of SFO do not coincide with  the temperature $T_{\rm FO}$, the chemical potentials $\mu_{B_{\rm FO}}$, $\mu_{S_{\rm FO}}$,
$\mu_{I3_{\rm FO}}$ and the volume $V_{\rm FO}$ which characterize the freeze-out of non-strange hadrons. The particle yields are given by the charge density $\rho^K_i$ in (1) and the corresponding
volume
at  FO and at  SFO.

 At the first glance a model with independent SFO contains four extra fitting  parameters for each  energy value compared to the traditional
approach (temperature, three chemical potentials and the volume at SFO instead of strangeness suppression/enhancement factor $\gamma_s$).
However, this  is not the case due to the conservation laws. Indeed, since the  entropy, the baryonic charge and the isospin third projection are conserved,  then the  parameters of FO and SFO are connected by the following equations
\begin{eqnarray}
s_{\rm FO} V_{\rm FO} = s_{\rm SFO} V_{\rm SFO} \,, \label{ent_cons}\\
\rho^B_{\rm FO} V_{\rm FO} = \rho^B_{\rm SFO} V_{\rm SFO} \,, \label{B_cons}\\
\rho^{I_3}_{\rm FO} V_{\rm FO} = \rho^{I_3}_{\rm SFO} V_{\rm SFO} \,, \label{I3_cons}
\end{eqnarray}
where the entropy density $s_{A} = \frac{\partial p}{\partial T}|_A$, the  density of baryonic charge 
$\rho^B_{A} = \frac{\partial p}{\partial \mu^B}|_A$ and the density of  the  isospin  third projection  
$\rho^{I3}_{A} = \frac{\partial p}{\partial \mu^{I3}}|_A$ are found from the usual thermodynamic identities 
at the SFO (A=SFO) or at FO (A=FO).

The effective volumes can be excluded,  if these equations are rewritten as
\begin{eqnarray}
\label{Eq:FO_SFO1}
\frac{s}{\rho^B} \biggl|_{\rm FO} = \frac{s}{\rho^B} \biggr|_{\rm SFO} \,,  \quad  \frac{\rho^B}{\rho^{I_3}} \biggl|_{\rm FO} = \frac{\rho^B}{\rho^{I_3}} \biggr|_{\rm SFO}
%
 \,.
\end{eqnarray}
Thus, the baryonic $\mu_{B_{\rm SFO}}$ and the  isospin third projection $\mu_{I3_{\rm SFO}}$ chemical potentials at SFO are now defined by Eqs. (\ref{Eq:FO_SFO1}). Note, that the strange chemical potentials $\mu_{S_{\rm FO}}$ and $\mu_{S_{\rm SFO}}$  are found from the condition of vanishing  net strangeness at FO and SFO,  respectively. Therefore, the concept of independent SFO leads to an appearance of one independently fitting parameter  $T_{\rm SFO}$.
Hence, the independent fitting parameters are the following: the baryonic chemical potential $\mu_B$, the chemical potential of the third projection of isospin $\mu_{I3}$,  the chemical freeze-out  temperature for strange hadrons $T_{\rm SFO}$, the chemical freeze-out  temperature for all non-strange hadrons $T_{\rm FO}$ and the $\gamma_s$ factor (i.e. 5 fitting parameters for each collision energy).

An inclusion  of  the width $\Gamma_i$ of  hadronic states  is an important element of the present model. It is due to the fact that the  thermodynamic properties of the hadronic system are sensitive to the width \cite{KABugaev:Horn2013, Bugaev13, Bugaev13new}. In order to account for the finite width of resonances we perform the usual modification of the thermal particle density $\phi_i$. Namely, we convolute the Boltzmann exponent under the integral over momentum with the normalized Breit-Wigner mass distribution. As a result, the modified thermal particle density of $i^{th}$ sort hadron acquires  the form
\begin{eqnarray}
\label{EqIII}
&&\int \exp\left(-\frac{\sqrt{k^2+m_i^2}}{T} \right)d^3k ~\rightarrow \nonumber \\
&& \frac{\int^{\infty}_{M_{0}} \frac{dx}{(x-m_{i})^{2}+\Gamma^{2}_{i}/4} \int \exp\left(-\frac{\sqrt{k^2+x^2}}{T} \right)d^3k }{\int^{\infty}_{M_{0}} \frac{dx}{(x-m_{i})^{2}+\Gamma^{2}_{i}/4}} \,.
\end{eqnarray}
Here $m_i$ denotes the  mean mass of  hadron and $M_0$ stands for the threshold in the dominant decay channel. The main advantages of this approximation is a simplicity of  its realization and a clear way to account for the finite  width of hadrons.
It is appropriate here to mention that one could use other prescriptions to account for the width of resonances. 
However,  in a recent work \cite{Bugaev13new} it was shown the Breit-Wigner prescription  (\ref{EqIII}) can provide  somewhat better (about 20\%) quality of the fit 
than the Gaussian  attenuation of resonance mass  and essentially better one compared to the case without accounting for the width. 
At the same time in \cite{Bugaev13new} it was found that  within the error bars such FO parameters as temperature, chemical potentials and $\gamma_s$ factor  are the same for different prescriptions of resonance width accounting.  Clearly, it is reasonable to expect that
other physically motivated ways to account for the resonance width should give similar results. 
Therefore,  in this work  we employed  the prescription   (\ref{EqIII}), which provides the better description of the data.

The observed hadronic multiplicities  contain the thermal and decay contributions. For example, a large part of pions is produced by the decays of heavier hadrons. Therefore, the total multiplicity is obtained as a sum of thermal and decay multiplicities, exactly as it is done in a conventional model. However, writing the formula for final particle densities, we have to take into account that the FO and SFO volumes  can be different:
\begin{eqnarray} \label{Eq:SFO_decays}
\frac{N^{fin}(X)}{V_{\rm FO}} = \sum_{Y \in FO} BR(Y \to X) \rho^{th}(Y) \nonumber \\
+
 \sum_{Y \in SFO} BR(Y \to X) \rho^{th}(Y) \frac{V_{\rm SFO}}{V_{\rm FO}} \,.
\end{eqnarray}
Here the first term on the right hand side is due to decays after FO whereas the second one accounts for the strange resonances decayed after SFO. The factor $V_{\rm SFO}/V_{\rm FO}$ can be replaced by $\rho^B_{\rm FO}/\rho^B_{\rm SFO}$ due to baryonic  charge conservation. $BR(Y \to X)$ denotes the branching ratio  of the Y-th hadron decay into the X-th hadron, with the definition $BR(X \to X)$ = 1 used  for the sake of convenience. The input parameters of the present model (masses $m_i$, widths $\Gamma_i$, degeneracies $g_i$ and branching  ratios of all strong decays) were taken from the particle tables of the thermodynamical code THERMUS \cite{THERMUS}.

\section{Fitting Procedures}

{\bf Data sets}. The present model is applied to fit the data. We take the ratios of particle multiplicities at midrapidity as the data points. In contrast to  fitting multiplicities themselves such an approach allows us  to cancel the possible experimental biases. In this paper we use the data set almost identical to Ref. \cite{Bugaev13}. At the AGS energies ($\sqrt{s_{NN}}=2.7-4.9$ AGeV or $E_{lab}=2-10.7$ AGeV) the data are available with a  good energy resolution above 2 AGeV. However, for the beam energies 2, 4, 6 and 8 AGeV only a few data points are available. They corresponds to the yields for pions \cite{AGS_pi1, AGS_pi2}, for protons \cite{AGS_p1,AGS_p2}, for kaons \cite{AGS_pi2} (except for 2 AGeV). The integrated over $4\pi$ data are also available for $\Lambda$ hyperons \cite{AGS_L} and for $\Xi^-$ hyperons (for 6 AGeV only) \cite{AGS_Kas}. However, as was argued in Ref. \cite{KABAndronic:05}, the data for $\Lambda$ and $\Xi^-$ should be recalculated for midrapidity. Therefore, instead of raw experimental data we used the corrected values from \cite{KABAndronic:05}. Next comes the data set at the highest AGS energy ($\sqrt{s_{NN}}=4.9$ AGeV or $E_{lab}=10.7$ AGeV). Similarly to \cite{KABugaev:Horn2013}, here  we analyzed  only  the  NA49  mid-rapidity data   \cite{KABNA49:17a,KABNA49:17b,KABNA49:17Ha,KABNA49:17Hb,KABNA49:17Hc,KABNA49:17phi}. Since  the RHIC high energy  data of different collaborations agree with each other, we  analyzed  the STAR results  for  $\sqrt{s_{NN}}= 9.2$ GeV \cite{KABstar:9.2}, $\sqrt{s_{NN}}= 62.4$ GeV \cite{KABstar:62a}, $\sqrt{s_{NN}}= 130$ GeV \cite{KABstar:130a,KABstar:130b,KABstar:130c,KABstar:200a} and $\sqrt{s_{NN}} =$ 200 GeV \cite{KABstar:200a,KABstar:200b,KABstar:200c}.

\begin{table}[t]
\begin{center}
\caption{The fit  results of different versions of  the HRGM are compared for 14 values of  the center of mass collision energies:  the column $\chi_1^2$ corresponds to a single FO model of
\cite{KABugaev:Horn2013}; the column $\chi_2^2$ is found for  the SFO with $\gamma_s=1$ \cite{Bugaev13};
 the column $\chi_3^2$   corresponds to the  SFO+$\gamma_s$ fit with added data points for $N_{rat} \le5$
 while   the column $\chi_4^2$  is obtained by the  direct SFO$\oplus \gamma_S$ fit.  $N_{rat}$ indicates the available number of independent hadronic ratios  at given  center of mass collision energy $\sqrt{s_{NN}}$. In the row Sum  we    list the sum of  $i$-th column, while  in the bottom row the number of degrees of freedom of each HRGM version is shown (for more details see the text).
}
\vspace*{4.4mm}
\begin{tabular}{cccccc}
\hline
~$\sqrt{s_{NN}}$~ &~ $\chi_1^2$~ & $N_{rat}$ & ~$\chi_2^2$ ~&~ $\chi_3^2$ ~&~ $\chi_4^2$~ \\
(GeV)& FO &  &  SFO & SFO+$\gamma_S$ & SFO$\oplus \gamma_S$ \\
 \hline
2.7 &   0.62 & 4      &  0.62  & 0.62  & $1.3 \cdot 10^{-5}$ \\ \hline
3.3 & 0.17   & 5  &     0.08   & 0.08  & $3.4 \cdot 10^{-9}$ \\ \hline
3.8 & 0.56   & 5  & 0.03   & 0.03  & 0.03 \\ \hline
4.3 & 0.35   & 5      & 0.26   & 0.26  & 0.21 \\ \hline
4.9 & 0.55   & 8      & 0.55   & 0.40  & 0.40 \\ \hline
6.3 & 7.91   & 9      & 2.88   & 2.45  & 2.45 \\ \hline
7.6 & 17.5   & 10 & 16.6   & 5.9   & 5.9  \\ \hline
8.8 & 7.9    & 11 & 7.85   & 7.56  & 7.56 \\ \hline
9.2 & 0.16   & 5  & 0.15   & 0.03  & $1.3 \cdot 10^{-7}$ \\ \hline
12  & 17.3   &10      & 11.9   & 9.57  & 9.57 \\ \hline
17  & 14.7   &13      & 7.39   & 7.38  & 7.38 \\ \hline
62.4    &  0.4   & 5  & 0.09   & 0.03  & 0.03 \\ \hline
130 &    5   & 11 & 4.62   & 4.32  & 4.32 \\ \hline
200 &  7.4   & 10 & 5.49   & 5.09  & 5.09\\ \hline \hline
Sum &  80.5 & 111 & 58.5 & 43.72 & 42.9 \\  \hline
Dof &  69 & N/A & 55 & 47 & 41 \\  \hline

\end{tabular}
\end{center}
\end{table}

{\bf Combined fit with SFO and $\gamma_{s}$ factor.}  A comprehensive data analysis \cite{Bugaev13}  performed recently for  two alternative  approaches, i.e  the first one with $\gamma_{s}$ as a free parameter and the second one with separate FO and SFO, showed  the advantages and disadvantages of both methods. Thus, the $\gamma_{s}$ fit provides one  with an opportunity to noticeably improve the Strangeness Horn description with $\chi^2/dof=3.3/14$, comparably to the previous result $\chi^2/dof=7.5/14$ \cite{KABugaev:Horn2013}, but there are only slight improvements of the ratios with strange baryons (global $\chi^2/dof: 1.16 \rightarrow 1.15$). The obtained results for the SFO approach demonstrate a high fit quality for the most problematic ratios for the HRGM, especially for   $\bar p/\pi^-$, $\bar \Lambda/\Lambda$,  $\bar \Xi^-/\Xi^-$ and  $\bar \Omega/\Omega$. Although the overall $\chi^2/dof \simeq 1.06$ is notably better than with the  $\gamma_{s}$ factor \cite{KABugaev:Horn2013, Bugaev13}, but the description of  the Horn's highest point  got worsen. These results  led  us to an idea to investigate the combination of these two approaches in order to get the high-quality Strangeness Horn description without spoiling the quality of other particle ratios.
{However, we immediately face a mathematical problem  to justify  such a combined fit because at six values of
the center of mass collision energies, namely  $\sqrt{s_{NN}}$ =2.7, 3.3, 3.8, 4.3, 9.2, 62.4 GeV,  the number of independent hadron yield ratios (4, 5, 5, 5, 5, 5, respectively) is equal or even smaller than the number of fitting parameters (see Table 1).
For these energies one, of course, can treat the experimental ratios as equations and can solve them, but, unfortunately,
the experimental ratios always have finite (and not small!) error bars. As a result,  solving the ratios as equations  with finite errors leads to rather  large region of chemical FO parameters which provide a vanishing value of  $\chi^2$ and, hence, it is hard to conclude what values are the most probable ones.
It seems that these difficulties prevented  the authors of a recent work  \cite{BugaevIndia2} to analyze the data at the collision energies $\sqrt{s_{NN}} \le $ 4.9   GeV within their version of SFO concept  \cite{BugaevIndia}.
}

\begin{figure}[t]
\begin{minipage}[h]{0.88\linewidth}
\center{\includegraphics[width=1.0\linewidth]{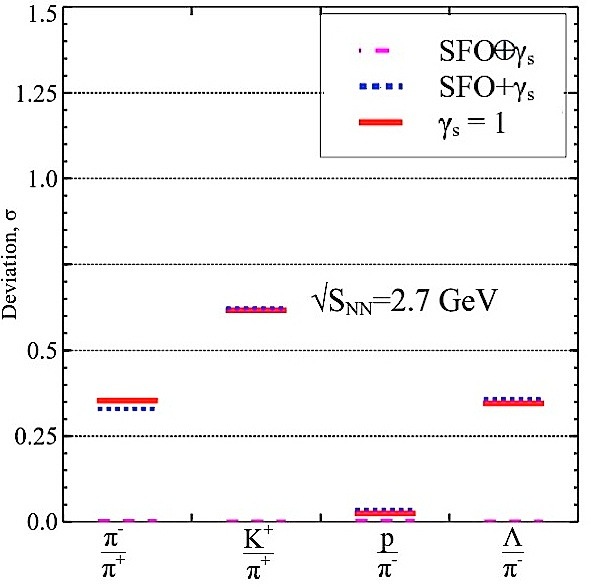}}
\end{minipage}

\begin{minipage}[h]{0.88\linewidth}
\center{\includegraphics[width=1.0\linewidth]{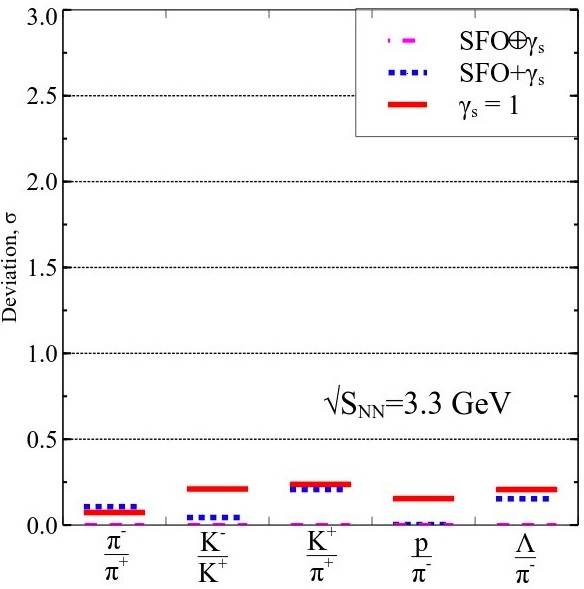}}
\end{minipage}
\caption{(Colour on-line)  Relative deviation of the theoretical description of ratios from the experimental value in units of the experimental error $\sigma$. Particle ratios vs. the modulus of relative deviation ($\frac{|r^{theor} - r^{exp}|}{\sigma^{exp}}$) for  $\sqrt{s_{NN}} = $ 2.7  and 3.3 GeV are shown. Solid lines correspond to the model with a single FO of all hadrons and $\gamma_s =1$, the dotted lines correspond to the model  SFO$+\gamma_s$ as explained in the text. The results of  the    SFO$ \oplus  \gamma_s$  model  are  indicated by the dashed lines.
}
\label{Fig:sagunI}
\end{figure}

{
Moreover, in some cases  the range of  chemical FO  parameters  obtained by such a fit, the SFO$\oplus \gamma_S$ hereafter,
is  located far away from the ones found by the well established fit procedures, i.e. by the single FO model without \cite{KABugaev:Horn2013} or with \cite{Bugaev13} the $\gamma_s$ fit  and  by  the SFO with $\gamma_s=1$ \cite{Bugaev13} which provide us with very good descriptions of the data.
Moreover, all these results are in a very good correspondence with each other.
Therefore, the combined SFO and $\gamma_S$ fit can be directly  performed  for the collision energies $\sqrt{s_{NN}} = $  4.9, 6.3, 7.6, 8.8, 12, 17, 130, 200 GeV only, while for other energies we have to seek for another minimization criterion.

Since the major task of the present work is to determine the  residual effect of the strangeness non-equilibrium on top of the SFO, then it would be reasonable to fix the parameters of the SFO and make the $\gamma_s$ fit. Unfortunately, in this case the number of degrees of freedom will be the same as for the SFO$\oplus \gamma_S$ fit, i.e. dof = 41,  and the resulting value of  $\chi^2/dof$ would not be better than  the one obtained within  the SFO$\oplus \gamma_S$ fit.  Hence, to avoid the above mentioned problems  we suggest  to modify the definition of $\chi^2$  for the six values of the collision energy
which have 5 or less independent ratios
\begin{eqnarray}\label{EqXI}
&&\chi^2 =  \sum_i\frac{(r_i^{theo}-r_i^{exp})^2}{\sigma^2_i} \nonumber \\
 &&+ \left[\frac{T_{\rm SFO} - T_{\rm SFO}(\gamma_s=1)}{\sigma_{T}^{\rm SFO}} \right]^2 \,,
\end{eqnarray}
where $r_i^{theo}$ and $r_i^{exp}$ are, respectively,  the  theoretical and the experimental  values of particle yields ratios, $\sigma_i$ stands for the corresponding experimental error and a  summation is performed over all  experimental points which are available at the considered  energy.  Here $\sigma_{T}^{\rm SFO}$ denotes the error of  the  SFO temperature $T_{\rm SFO}(\gamma_s=1)$ which is found for each problematic energy by  the SFO fit with $\gamma_s=1$, while the chemical FO temperature of strange particles $T_{\rm SFO}$  is the fitting parameter.

In other words,  for each energies  corresponding to a set of problematic ratios we suggest  to consider the SFO temperature $T_{\rm SFO}(\gamma_s=1)$, as an additional datum  to be fitted within the combined SFO+$\gamma_s$ approach, while for other collision energies we used the standard  definition $\chi^2 =  \sum_i\frac{(r_i^{theo}-r_i^{exp})^2}{\sigma^2_i}$ for  the combined fit. In order to distinguish this approach from the SFO$\oplus \gamma_S$ fit we refer to  it as  the SFO+$\gamma_s$ fit.  Such a reformulation of  minimization criterion for $\sqrt{s_{NN}}$ =2.7, 3.3, 3.8, 4.3, 9.2, 62.4 GeV allows us to avoid the mathematical problems of  the combined   SFO$\oplus\gamma_s$ fit  and to  simultaneously keep  the temperature of strange particles $T_{\rm SFO}$
not far away from the SFO temperature $T_{\rm SFO}$.  Originally, for $\sqrt{s_{NN}}$ =2.7 we added two data points  into the $\chi^2 $ definition (\ref{EqXI})  in order to have 6 data for 5 fitting parameters, but then we found that adding one data point  is sufficient, since it resolves the problem.

\section{Main Results}

The results of the SFO+$\gamma_s$, SFO$\oplus \gamma_S$ and SFO fits are compared for $\sqrt{s_{NN}}$ =2.7 and 3.3 GeV in Fig. \ref{Fig:sagunI} (for more details see also Table I). As one can see from this  figure
the SFO description is already very good ($\chi^2 \equiv \chi^2_2 \simeq 0.62$ in Table I) and, hence, the additional parameter $\gamma_s$  cannot improve it (compare $\chi^2_3$ and  $\chi^2_2$ in Table I), if the number of data points is equal or large than the number of fitting parameters.  Thus, the mathematically justified
SFO+$\gamma_s$ fitting procedure does not improve the description quality compared to the SFO fit at these collision energies
and, hence, we find that $\gamma_s \simeq 1$ within the error bars.
Moreover, for a completeness  we used another way of fitting:  first we determined the parameters of  two chemical freeze-out within the SFO model with $\gamma_s=1$ (see the $\chi^2_2$ column  in Table I), fixed the found parameters and then we  performed the fitting of  the $\gamma_s$ parameter. It is remarkable that in this way we did not get any improvement of the fit quality compared to the SFO+$\gamma_s$ fit and got the same
freeze out temperatures and chemical potentials as in the latter case not only  for the problematic data points,
but within the error bars we found the same results for all other energies of collision.

The main results  for the SFO$\oplus \gamma_S$ and SFO+$\gamma_s$ fits found here  are as follows.
Due to the problems discussed above the SFO$\oplus \gamma_S$ model does not allow us to locate
the narrow region of the chemical FO parameters for the collision energies $\sqrt{s_{NN}}$ =2.7, 3.3 and  9.2 GeV, while for  the energies $\sqrt{s_{NN}}$ = 3.8, 4.3 and 62.4 GeV we did not find the solutions of five equations for five variables and, hence, were able to perform the usual minimization of $\chi^2$.

The SFO$\oplus \gamma_S$ model  gives  $\chi_4^2/dof$ = 42.96/41 $\simeq$ 1.05, which is only a very slight improvement compared to the previously obtained  results for the SFO model $\chi^2_2/dof$ =58.5/55 $\simeq$ 1.06. The redefinition of the  $\chi^2$ criterion (\ref{EqXI}) allows us to avoid the mathematical problems
within the SFO+$\gamma_s$ model and to sizably  reduce the  $\chi^2$ value per degree of freedom  to $\chi_3^2/dof$ = 43.72/47 $\simeq$ 0.93.  Moreover, for the problematic data at the collision energies $\sqrt{s_{NN}}$ = 3.8, 4.3 and 62.4 GeV  within the SFO+$\gamma_s$ model we obtained practically the same chemical  FO parameters and the same quality of the fit (compare the values of $\chi_3^2$ and $\chi_4^2$ in Table I for these energies), as for the SFO$\oplus \gamma_S$ model, including the main conclusion that  $\gamma_s \simeq 1$ (see the upper panel of Fig. \ref{Fig:sagunII}). Such a result provides an additional justification for  the   $\chi^2$ criterion redefinition (\ref{EqXI}).

Nevertheless,  as one can see from Table I compared to the SFO model with $\gamma_s=1$ (see the column with $\chi^2_2$) the main reduction of  $\chi^2$
achieved by the $\gamma_S$ parameter corresponds to $\sqrt{s_{NN}} = $ 7.6 GeV, i.e this is exactly where the Strangeness Horn peak is located.
Morevover,
we found that the fitting results can be separated into two distinct groups: those, where $\chi^2 > 1$ and where $\chi^2 < 1$ for any of our fits. It is remarkable that  neither SFO, nor $\gamma_s$ fits  do not move any of the points
of one group to another group. If for a certain collision energy the inequality  $\chi^2 > 1$ occurred, then   it always holds after any of our efforts.  The results obtained for the SFO+$\gamma_s$ are shown in Figs.
\ref{Fig:sagunII}--\ref{Fig:sagunVIII}.
Note   that compared to the SFO model with $\gamma_s=1$ \cite{Bugaev13}  the value of $\chi^2$ itself for  the  SFO$+\gamma_S$ fit, not divided by number of degrees of freedom,  has improved notably, although the deviation of the $\gamma_s$ factor from
1  does not exceed 20 \%  even for the topmost point of the Strangeness Horn (see the upper panel of Fig. \ref{Fig:sagunII}).
Note that our results on the  SFO$+\gamma_S$ model are very similar to the SFO model of Ref. \cite{BugaevIndia2} (just compare our Fig. 3 with Fig. 4 in \cite{BugaevIndia2}), although at the energies  $\sqrt{s_{NN}} =130 $ GeV and  $\sqrt{s_{NN}} = 200 $ GeV we find that the temperature of FO is slightly higher than the  temperature of SFO, while in  \cite{BugaevIndia2} the situation is opposite.  Two possible reasons for such a difference is that in  Ref. \cite{BugaevIndia2} the conservation laws (5)-(7) are  ignored and their treatment is based on the ideal gas picture.  As a result the fit quality achieved in
\cite{BugaevIndia2} is essentially lower (see Fig. 5 in there) compared to the present work.

It is remarkable that the present rather sophisticated fit of the hadronic multiplicities confirms the recent finding on the non-smooth behavior of the function $T_{\rm FO}(\mu_{B}^{\rm FO})$  reported in \cite{Bugaev13new} for the same  hard-core radius of hadrons $R=0.3$ fm. A similar  change of the slope of   $T_{\rm SFO}(\mu_{B}^{\rm SFO})$ occurring at the
collision energy $\sqrt{s_{NN}} \simeq $ 4 GeV is a new result shown in Fig. \ref{Fig:sagunII}.
Following the work   \cite{Bugaev13new}, we parameterize $T_{\rm FO}(\sqrt{s_{NN}})$, $T_{\rm SFO}(\sqrt{s_{NN}})$,
$\mu_{B}^{\rm FO}(\sqrt{s_{NN}})$ and $\mu_{B}^{\rm SFO}(\sqrt{s_{NN}})$ as
\begin{eqnarray}\label{EqXII}
T &=& (T_1 + T_2 \sqrt{s_{NN}}) \cdot c_+ ( \sqrt{s_{NN}}, 4.0, 0.1) +  \nonumber \\
&+&(T_3/ \sqrt{s_{NN}} + T_4) \cdot c_- ( \sqrt{s_{NN}}, 4.0, 0.1)\, , \\
\mu_B &=& \frac{A}{1+B\sqrt{s_{NN}}}\, ,
\label{EqXIII}
\end{eqnarray}
where  $c_\pm (x,a,b)$ are the  sigmoid functions
\begin{eqnarray}\label{EqXIV}
c_+(x,a,b) &=& \frac{1}{1+e^{(x-a)/b}} = \frac{1}{2}\left(1-\tanh{\frac{x-a}{2b}}\right)\, , \\
c_-(x,a,b) &=& \frac{1}{1+e^{(a-x)/b}} = \frac{1}{2}\left(1+\tanh{\frac{x-a}{2b}}\right)\, .
\label{EqXV}
\end{eqnarray}
The main reason to employ the parameterizations  (\ref{EqXII})-(\ref{EqXV}) is a  drastic change of  the
$\sqrt{s_{NN}}$ dependence of  the function $T_{\rm FO}(\sqrt{s_{NN}})$ in the narrow region $\sqrt{s_{NN}}\simeq 4.3 - 4.9$ GeV found in \cite{Bugaev13new}. The upper panel of  Fig. \ref{Fig:sagunIII}
confirms  that a similar behavior  of  $T_{\rm FO}(\sqrt{s_{NN}})$ and  $T_{\rm SFO}(\sqrt{s_{NN}})$ exists
within the SFO+$\gamma_s$ model, although the switch value of the collision energy $\sqrt{s_{NN}} \simeq $ 4 GeV is about ten percents lower than the one found for the single FO model of work \cite{Bugaev13new}. The resulting curves (\ref{EqXII})  and (\ref{EqXIII})  for the SFO+$\gamma_s$ model are shown in Fig. \ref{Fig:sagunIII}, whereas the corresponding parameters are give in Table 2.
Using curves (\ref{EqXII})  and (\ref{EqXIII}) we obtained the analytic expressions  for the functions
$T_{\rm FO}(\mu_{B}^{\rm FO})$ and  $T_{\rm SFO}(\mu_{B}^{\rm SFO})$, which are shown in the lower panel of  Fig.
\ref{Fig:sagunII}.

For a comparison in Figs. \ref{Fig:sagunII} and \ref{Fig:sagunIII} we depict the parameterization
of the chemical freeze out temperature ($\sqrt{s_{NN}}$ is given in GeVs)
\begin{eqnarray}\label{EqXN}
T^{\rm FO}[MeV]  &= & \frac{T^{lim}}{1 + \exp \left[  2.60 - \ln(\sqrt{s_{NN}})/0.45    \right] } = \nonumber \\
 &= & T^{lim} \, c_-(\ln(\sqrt{s_{NN}}), 1.17, 0.45) \,,
\label{EqXVI}
\end{eqnarray}
which together with the parameterization (\ref{EqXIII}) was suggested in  \cite{KABAndronic:05}.
The parameters   $T^{lim} = 164 $ MeV,  $A = 1303$\, MeV, $B = 0.286 $\, GeV$^{-1}$
were found in   \cite{KABAndronic:09}. Although they have close values to $T_4$, $A $ and $B$ listed in Table,
only the curves $\mu_B (\sqrt{s_{NN}})$ found here and in  \cite{KABAndronic:09} look similar (see the lower panel of  Fig. \ref{Fig:sagunIII}).
  From the upper  panel of  Fig.  \ref{Fig:sagunIII} one can see that
the curves  (\ref{EqXII}) and (\ref{EqXVI}) have  rather different behavior at low and intermediate values of collision energy. As a result the functions   $T_{\rm FO}(\mu_{B}^{\rm FO})$  found here and in  \cite{KABAndronic:09} have different shapes as one can see from the lower panel of  Fig.    \ref{Fig:sagunII}. We have to note that our efforts to reasonably describe the FO and SFO temperatures by the  parameterization (\ref{EqXVI})  were not successful  and the corresponding values of $\chi^2/dof$  were almost one order of magnitude  larger than the ones
given in Table  2 for  Eq.  (\ref{EqXII}).
Clearly, the found parameterizations should be considered as the predictions for  the chemical FO and SFO characteristics which can be experimentally  tested at the accelerators  FAIR (GSI, Darmstadt) and NICA (JINR, Dubna).

\begin{figure}[!]
\center{\includegraphics[width=77mm]{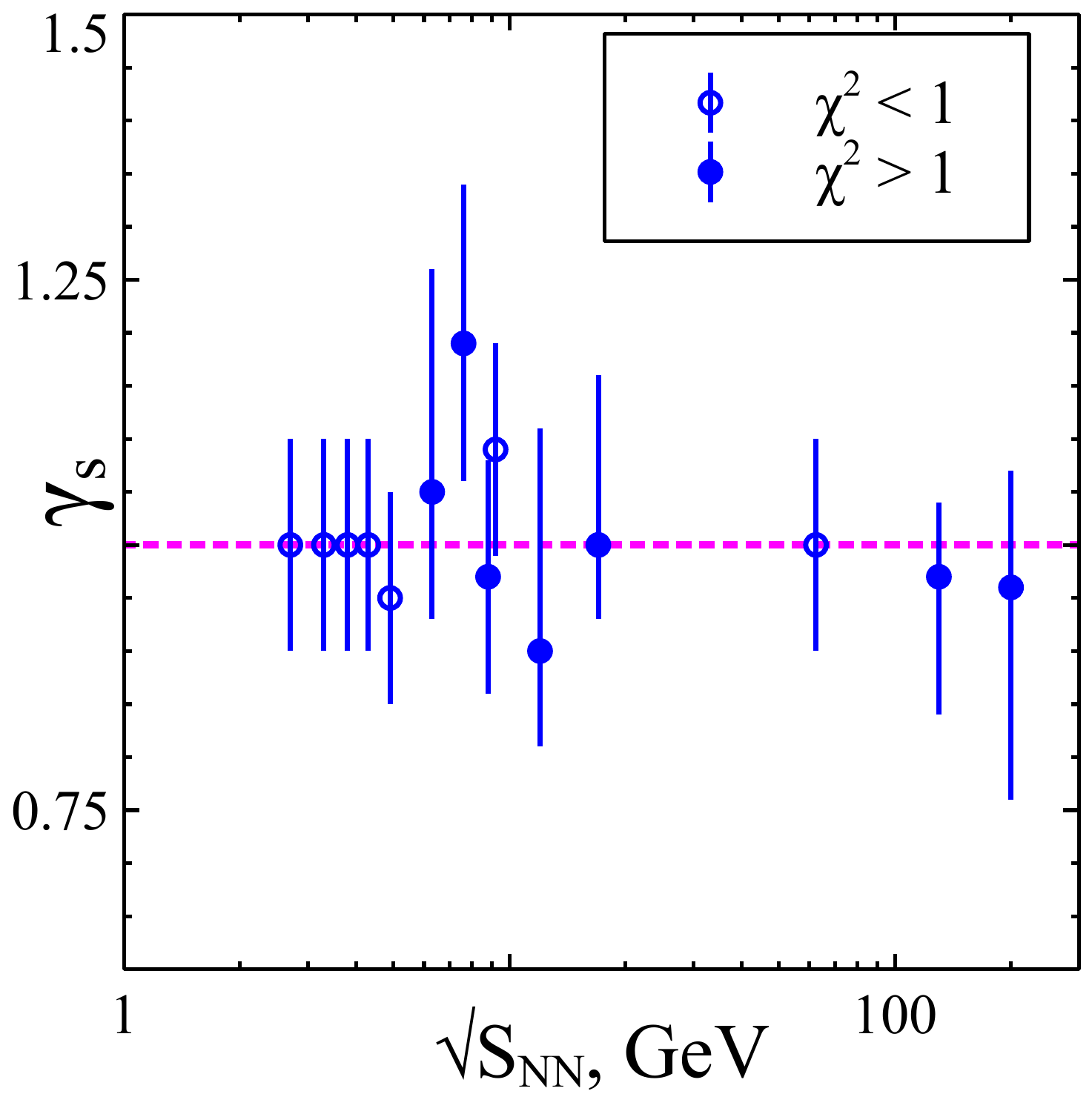}

\vspace*{2.2mm}
 \includegraphics[width=77mm]{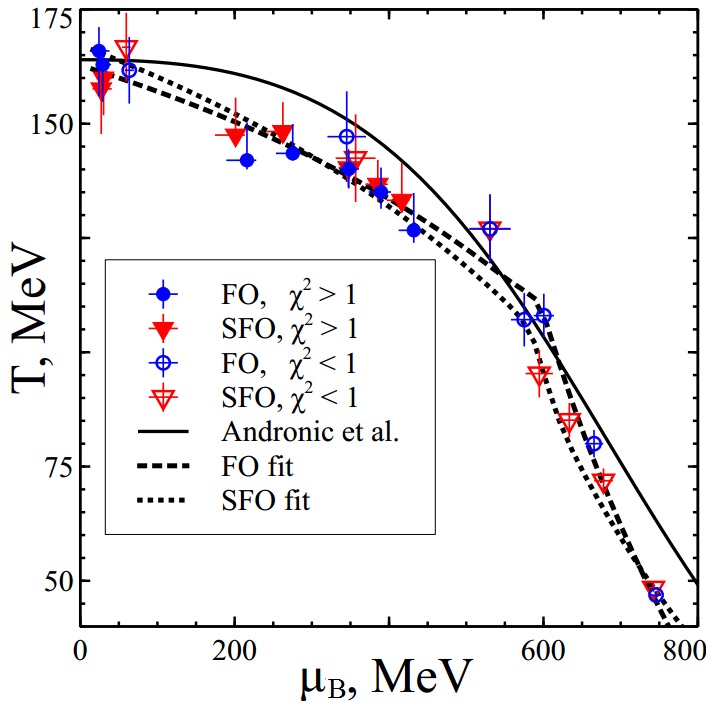}}
\caption{(Colour on-line)
{\bf Upper panel:} $\sqrt{s_{NN}}$ dependence of the $\gamma_s$ factor within  the SFO$+\gamma_S$ model with two freeze-outs and the  $\gamma_s$ fit.
{\bf  Lower panel:}
Chemical freeze-outs parameters found within  the  SFO$+\gamma_{s}$ model. Baryonic chemical potential dependence of the chemical freeze-out temperature for the strange hadrons (SFO points are marked with triangles) and for  the non-strange ones (FO points are marked with circles). The pairs of nearest  points are connected  by the  isentrops $s/\rho_{B}=const$, on which the FO and the SFO points are located.}
\label{Fig:sagunII}
\end{figure}

\begin{figure}[!]
\begin{minipage}[h]{0.95\linewidth}
\center{\includegraphics[width=.95\linewidth]{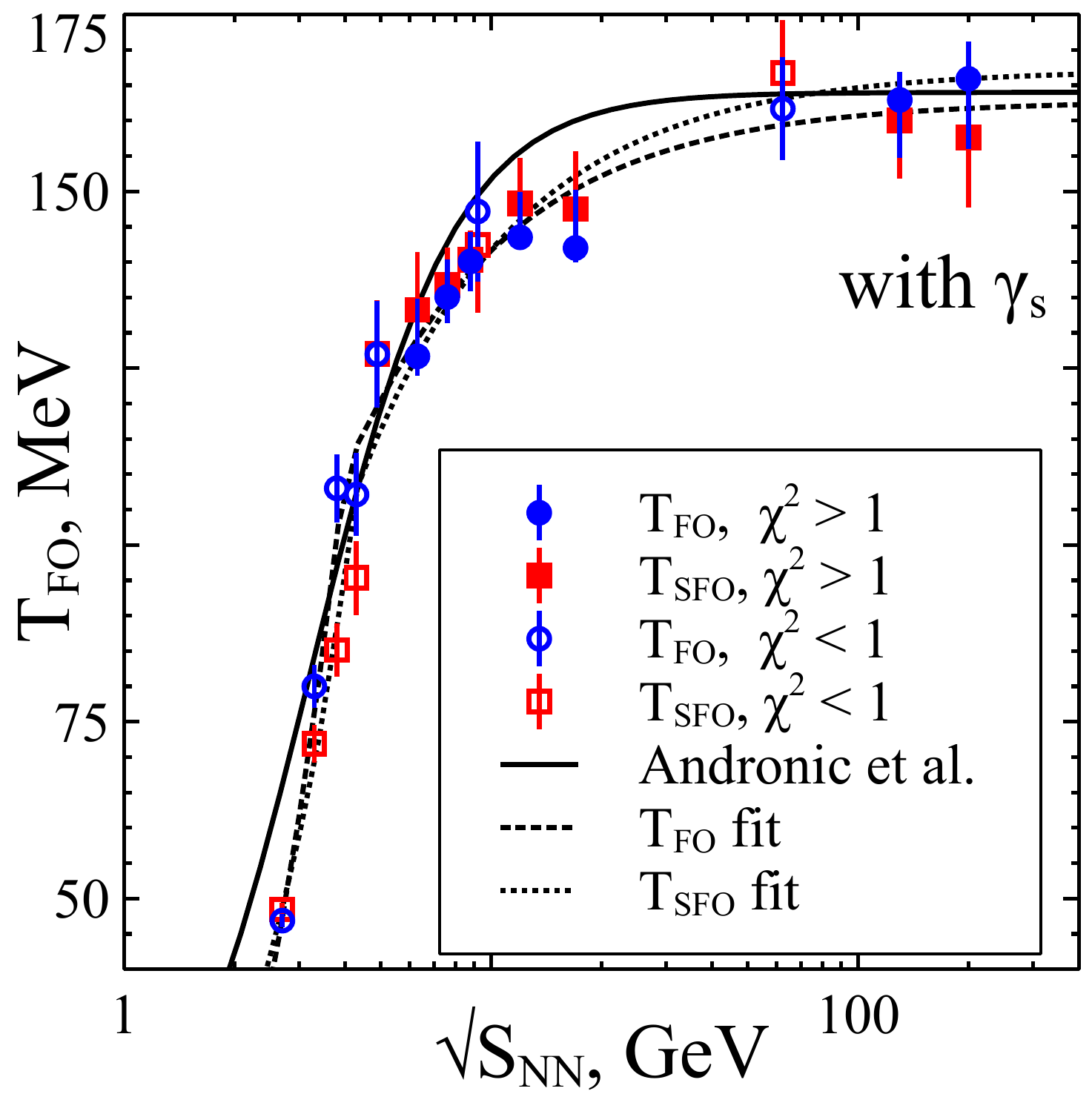}}
\end{minipage}

\begin{minipage}[h]{0.93\linewidth}
\center{~\includegraphics[width=1.\linewidth]{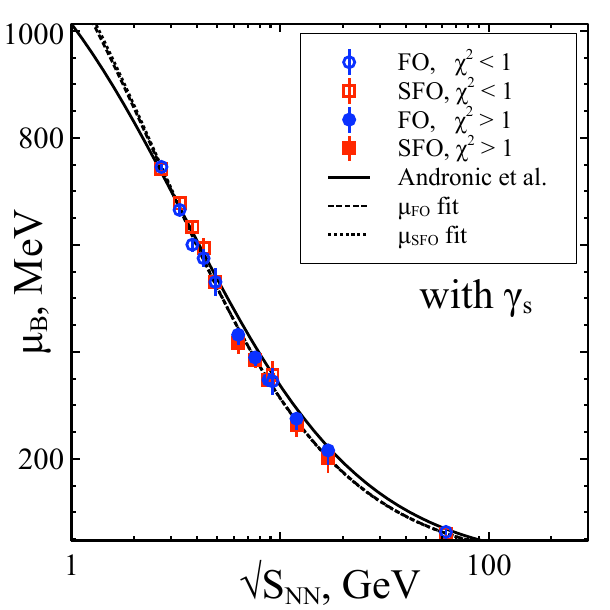}}
\end{minipage}
\caption{(Colour on-line) The behavior of the SFO$+\gamma_S$ model parameters: the chemical  freeze-out temperatures $T_{\rm FO}$ and $T_{\rm SFO}$ vs. $\sqrt{s_{NN}}$ (upper panel) and the freeze-out baryonic chemical potentials $\mu_{B}^{\rm FO}$ and $\mu_{B}^{\rm SFO}$ vs. $\sqrt{s_{NN}}$  (lower panel).
}
\label{Fig:sagunIII}
\end{figure}

\begin{figure}[!]
\vspace*{-0.7mm}
\center{\includegraphics[width=79mm]{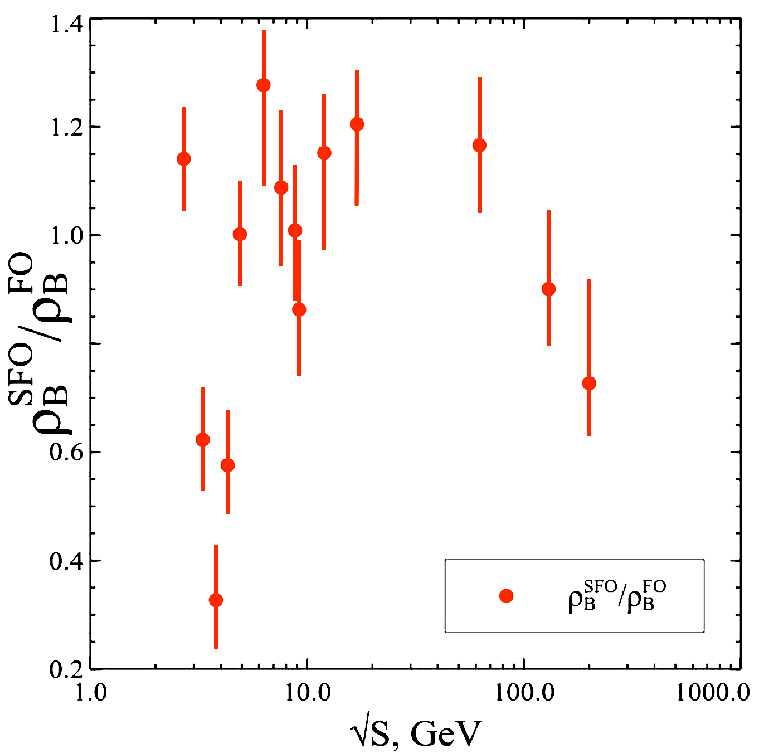}}
\caption{(Colour on-line)
 $\sqrt{s_{NN}}$ dependence of the ratio of  baryonic charge densities at SFO and at FO  within  the SFO$+\gamma_S$ model. Since the baryonic charge is conserved, then such a ratio coincides with the inverse ratio of  corresponding freeze-out volumes, i.e. with $V_{\rm FO}/V_{\rm SFO}$.
}
\label{Fig:sagunIIIb}
\end{figure}


 \begin{table}[!]
\begin{center}
\caption{The parameters of  Eqs.  (\ref{EqXII})  and (\ref{EqXIII})  found from  fitting the
values of chemical freeze out parameters of the SFO+$\gamma_s$ model.
}
\vspace*{4.4mm}
\begin{tabular}{lll}
\hline
         & FO     & SFO     \\ \hline
$T_1$ ({\rm MeV})    & -82.7  & -43.9   \\ \hline
$T_2$  ({\rm MeV})   & 48.1   & 34.2    \\ \hline
$T_3$  ({\rm MeV})   & -211.8 & -254.8  \\ \hline
$T_4$ ({\rm MeV})    & 162.8  & 167.2   \\ \hline
$\chi^2/dof$ fit Eq. (\ref{EqXII})  & 16/9    & 14.5/9    \\ \hline
$A$  ({\rm MeV})     & 1501   & 1525    \\ \hline
$B   ({\rm GeV}^{-1}) $   & 0.38   & 0.39    \\ \hline
$\chi^2/dof$ fit Eq. (\ref{EqXIII})  & 3.4/12    & 8.1/12     \\ \hline
\end{tabular}

\end{center}
\end{table}

{
In Fig. \ref{Fig:sagunIIIb} we show the ratio of  baryonic charge densities at SFO and at FO  within  the SFO$+\gamma_S$ model,
which  coincides with the inverse ratio of  corresponding freeze-out volumes $V_{\rm FO}/V_{\rm SFO}$ due to the baryonic charge conservation. From this figure one can see that  visually small difference of  temperatures and baryonic chemical potentials at SFO and FO leads, nevertheless, to quite sizable difference of  other thermodynamic quantities. As one can see from the upper panel of Fig. \ref{Fig:sagunIV}, this is also true for the pressure existing  at SFO and at FO. 
}

Also we confirm the existence of  irregularities   in the FO pressure  found earlier \cite{Bugaev13new}.
 Similar irregularities we find for  the SFO pressure and
  for  the effective number of degrees of freedom for SFO, $p^{\rm SFO}/(T^{\rm SFO})^4$ , i.e. for the ratio of SFO pressure to the fourth power of the SFO temperature.  From the upper panel of Fig. \ref{Fig:sagunIV} one can see that the largest increase of  SFO pressure per  increase of  the center of mass energy of collision occurs at  $\sqrt{s_{NN}}= 4.3 -4.9$ GeV.
 In other words, for about 14\% increase of   $\sqrt{s_{NN}}$ the SFO  pressure increases in 5.3 times and
 the ratio
 $p^{\rm SFO}/(T^{\rm SFO})^4$ increases  on about 65 \%.  According to the recent work \cite{Bugaev:2014uha}
 these and other irregularities observed at chemical FO  \cite{Bugaev13new} are signaling about the formation of the mixed quark-gluon-hadron phase and such an  explanation can be experimentally  verified  in a few years
 at FAIR  and NICA.

One more important finding of the present work can be seen from a comparison of the upper and lower panels of
Fig. \ref{Fig:sagunIV}. Note that in contrast to the temperature or baryonic chemical potential,  the pressure
allows one easier to distinguish the SFO from FO. Moreover, comparing the squares in the upper  and lower  panels of
Fig. \ref{Fig:sagunIV} one immediately concludes that the model of a single chemical FO   \cite{Bugaev13new}
with the same value of hard-core radius for all hadrons describes just the SFO for all values of  $\sqrt{s_{NN}}$
below 62.4 GeV.  The same is true for a single FO model  \cite{KABugaev:Horn2013} with different hard-core radii discussed above.
This peculiar result can be easily  understood, if one recalls that  for the most values  of collision energy the  number of  ratios involving  strange  hadrons is essentially larger than the number of  ratios with non-strange particles.

\begin{figure}[!]
\center{\includegraphics[width=80mm]{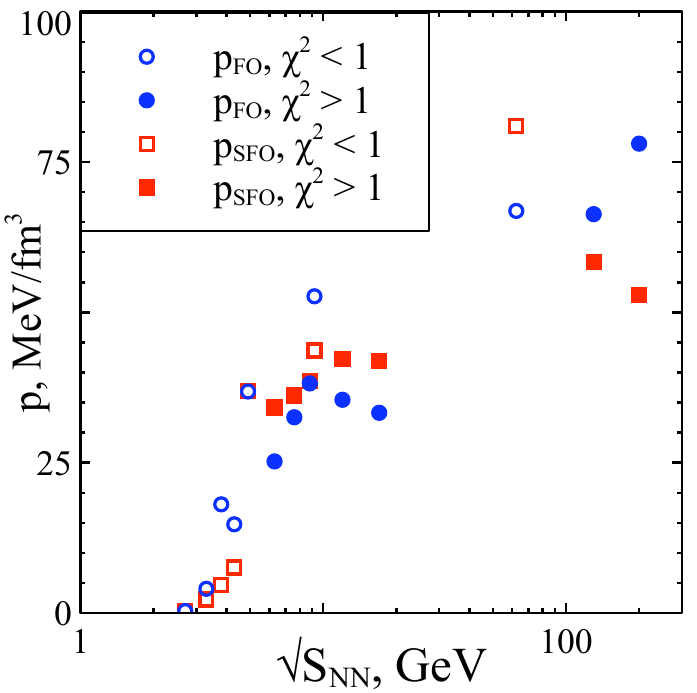}

 \includegraphics[width=79mm]{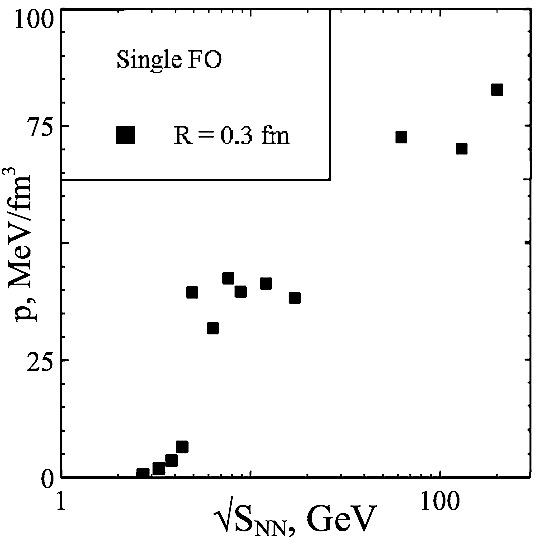}}
\caption{(Colour on-line)  $\sqrt{s_{NN}}$  dependences of  the pressure at FO and SFO points found within the  SFO+$\gamma_s$model (upper panel)  and within a single FO model for the same hard-core radius $R=0.3$ fm of all hadrons \cite{Bugaev13new}.
}
\label{Fig:sagunIV}
\end{figure}

 At the same time   a single chemical FO  model reproduces the FO pressure of the SFO+$\gamma_s$ model  only at  $\sqrt{s_{NN}} \ge 62.4$  GeV.  The corresponding reasons we will discuss in the next section, while here
we mention that at high RHIC energies the fit quality of all models, including the single FO one, is rather high
as one can see from Table I. The main part of $\chi^2$ at these energies  is formed by
poor description   of   $K$-mesons  at $\sqrt{s_{NN}} = 62.4$  GeV,   $\Lambda$ hyperon  at $\sqrt{s_{NN}} = 130$  GeV
and    $\Omega^\pm$   hyperons  at $\sqrt{s_{NN}} = 200$  GeV.

Such a comparison of the  single FO model and  the SFO+$\gamma_s$ model  pressures allows us to explain a cause of
why in previous thorough  analysis of the particle ratios within  the realistic single FO model  with the same hard-core radius of  hadrons  the main conclusion was that there is no deviation of strange particles from chemical equilibrium for the mid-rapidity data. In other words, it is possible to naturally explain the reason of  why  in  \cite{KABAndronic:05} it was found that  within the error bars $\gamma_s \simeq 1$. Our direct comparison
of the SFO+$\gamma_s$ model
for the mid-rapidity data
shows that the single FO models with the same or different hard-core hadronic radii reproduce the SFO  pressure
with $\gamma_s \simeq 1$ almost at all collision energies, except for  $\sqrt{s_{NN}} =7.6$ GeV and for  $\sqrt{s_{NN}} \ge  62.4$  GeV. In the former case one finds that $\gamma_s \simeq 1.19$, while in the latter case  $\gamma_s \simeq 1$, but, as we discussed  above,
for high collision  energies the pressure at FO found within  the SFO+$\gamma_s$ model   reproduces
the single FO model pressure. Therefore,
the main reason of why the chemical non-equilibrium  of strange charge was not found within the single FO models is  that  the fitting procedure mainly described the ratios involving the strange particles and, hence, it would have been   more appropriate to consider  the chemical non-equilibrium of non-strange hadrons.
}

\section{Results for Particle Ratios}

\begin{figure}[!]
\begin{minipage}[h]{0.88\linewidth}
\center{\includegraphics[width=1.0\linewidth]{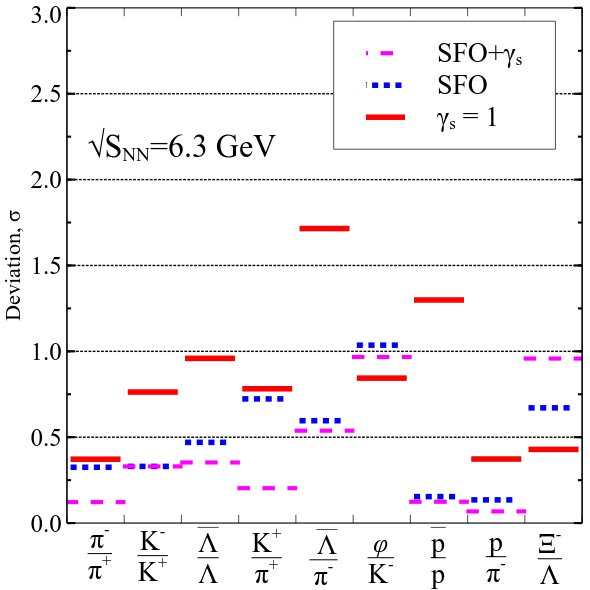}}
\end{minipage}
\begin{minipage}[h]{0.88\linewidth}
\center{\includegraphics[width=1.0\linewidth]{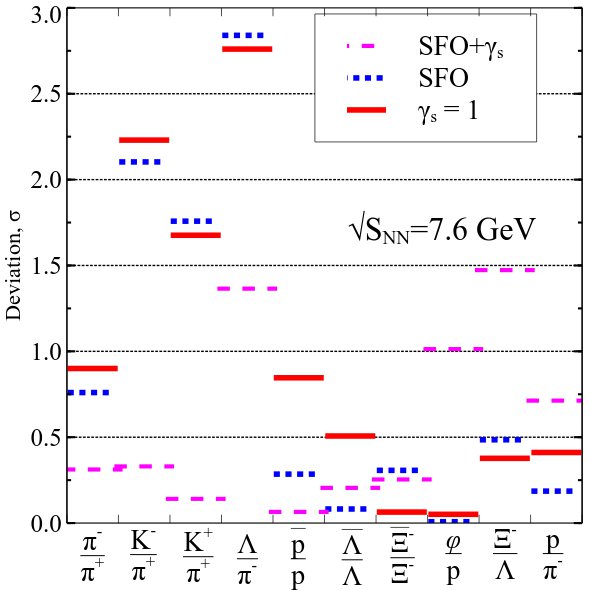}}
\end{minipage}
\caption{(Colour on-line) Relative deviation of the theoretical description of ratios from the experimental value in units of the experimental error $\sigma$. Particle ratios vs. the modulus of relative deviation ($\frac{|r^{theor} - r^{exp}|}{\sigma^{exp}}$) for  $\sqrt{s_{NN}} = $ 6.3 and 7.6 GeV are shown. Solid lines correspond to the model with a single FO of all hadrons and $\gamma_s =1$, the  dotted lines correspond to the model with SFO. The results of a   model with  a combined  fit with SFO and  $\gamma_s$    are  indicated by the  dashed lines.
}
\label{Fig:sagunV}
\end{figure}

\begin{figure}[h]
\begin{minipage}[h]{0.88\linewidth}
\center{\includegraphics[width=1.0\linewidth]{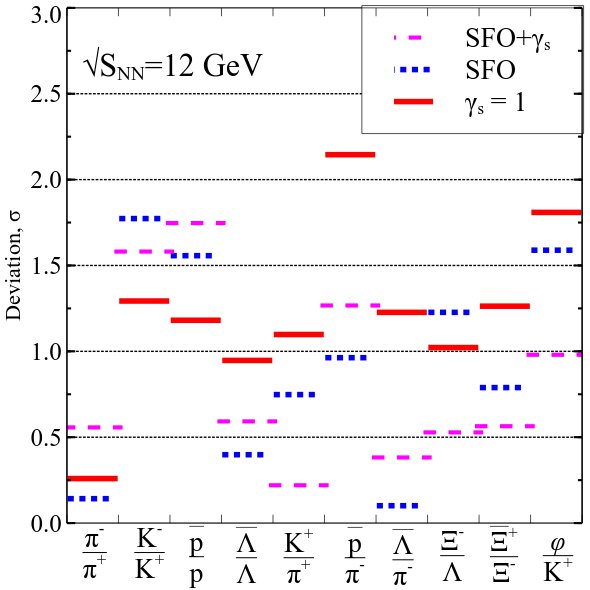}}
\end{minipage}

\begin{minipage}[h]{0.88\linewidth}
\center{\includegraphics[width=1.0\linewidth]{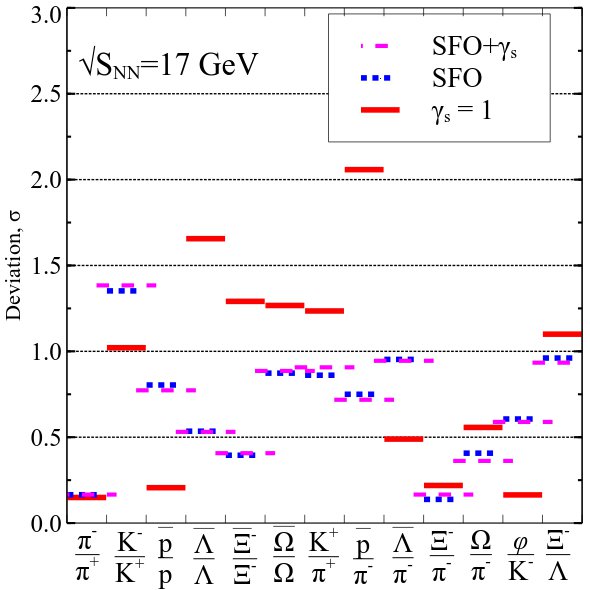}}
\end{minipage}
\caption{(Colour on-line) Same as in Fig. \ref{Fig:sagunV}, but for $\sqrt{s_{NN}} = $  12 and 17 GeV.
}
\label{Fig:sagunVb}
\end{figure}

The findings discussed above   motivate us to study in some details what ratios and at what energies  are  improved.
The most significant improvements correspond to the collision energies  $\sqrt{s_{NN}} = $ 6.3, 7.6, and 12 GeV, that are plotted in Figs. \ref{Fig:sagunV} and \ref{Fig:sagunVb}. Figs.  \ref{Fig:sagunV} and \ref{Fig:sagunVb} demonstrate very high fit quality, especially for such  traditionally  problematic ratios as $K^+/\pi^+$, $\pi^-/\pi^+$, $\bar \Lambda/\pi^-$ and $\varphi/K^+$,  which is achieved within  the SFO$+\gamma_S$ model  compared
to the single FO model and  the SFO one. For instance, for  $\sqrt{s_{NN}} = $ 7.6 GeV  seven ratios out of ten  are improved, while for other energies the improvements are less significant. On the contrary,  the particle ratios measured at  $\sqrt{s_{NN}} = $ 17  GeV (see Fig.  \ref{Fig:sagunVb})
are improved within the SFO model, while   the SFO$+\gamma_S$  fit   practically does not lead to any significant improvement compared to the SFO model.

\begin{figure}[h]
\center{\includegraphics[width=77mm]{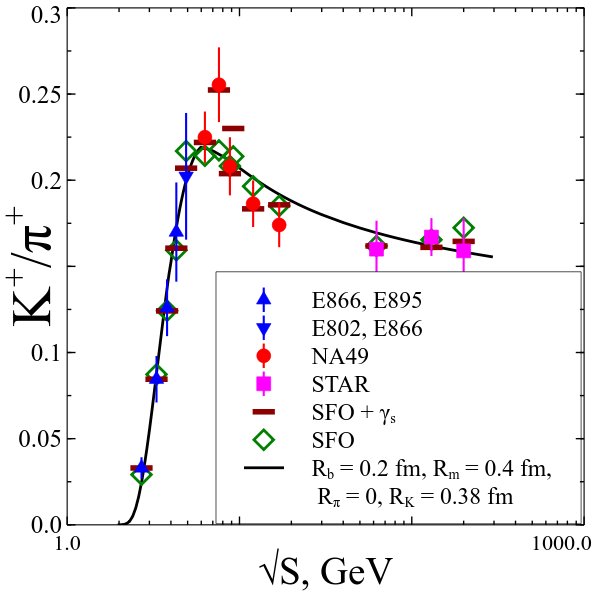}}
\caption{(Colour on-line)  $\sqrt{s_{NN}}$  dependences of  $K^+/\pi^+$ ratio. The solid line corresponds to the results of \cite{KABugaev:Horn2013}. Horizontal bars correspond to the present  model with SFO+$\gamma_s$ fit, while the  diamonds  correspond  to the  results  previously obtained for  SFO \cite{Bugaev13}.
}
\label{Fig:sagunVI}
\end{figure}

\begin{figure}[!]
\begin{minipage}[h]{0.88\linewidth}
\center{\includegraphics[width=1.0\linewidth]{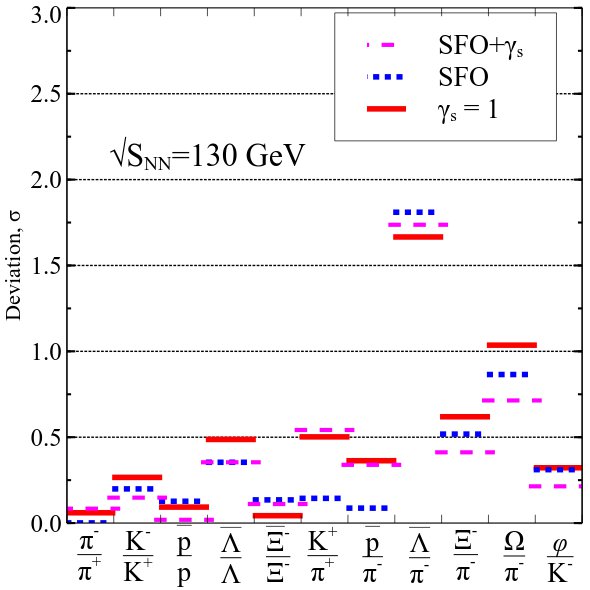}}
\end{minipage}

\begin{minipage}[h]{0.88\linewidth}
\center{\includegraphics[width=1.0\linewidth]{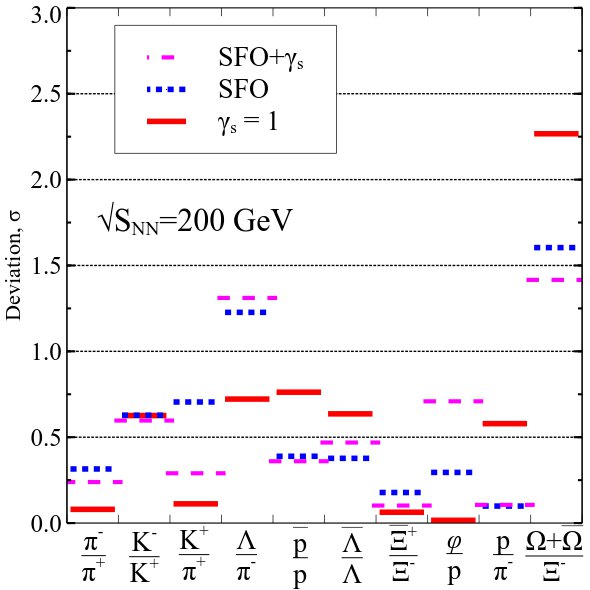}}
\end{minipage}
\caption{(Colour on-line) Same as in Fig. \ref{Fig:sagunV}, but  for  $\sqrt{s_{NN}} = $ 130 and 200  GeV.
}
\label{Fig:sagunVII}
\end{figure}

\begin{figure}[!]
\begin{minipage}[h]{0.88\linewidth}
\center{\includegraphics[width=1.0\linewidth]{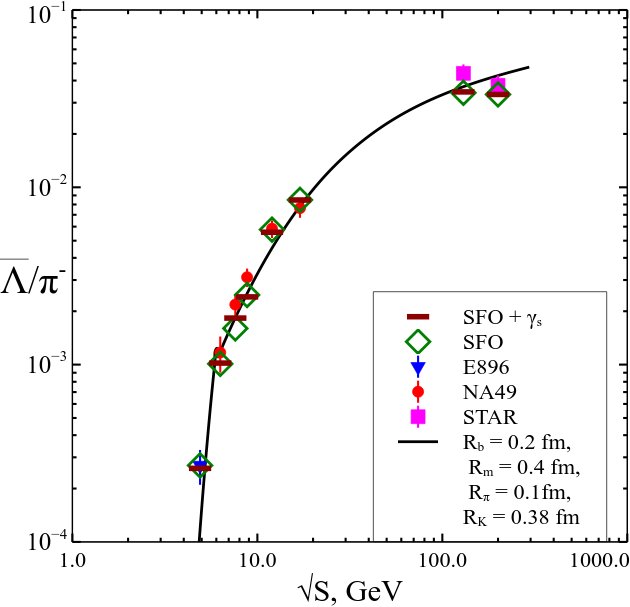}}
\end{minipage}
\hfill
\begin{minipage}[h]{0.88\linewidth}
\center{\includegraphics[width=1.0\linewidth]{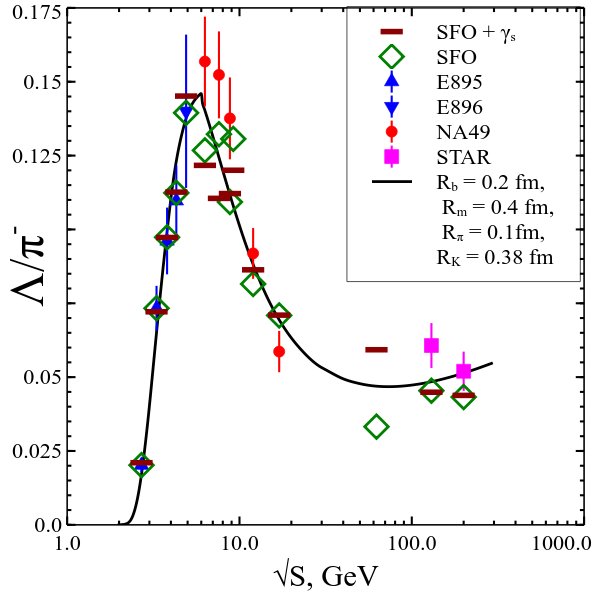}}
\end{minipage}
\caption{(Colour on-line)  $\sqrt{s_{NN}}$  dependences of  $\bar \Lambda/\pi^-$ (upper  panel) and $\Lambda/\pi^-$ (lower panel) ratios. The solid line correspond to the results of  a single FO model  \cite{KABugaev:Horn2013}.  Horizontal bars correspond to SFO+$\gamma_s$ model, while the  diamonds  correspond  to the previously obtained results  for the  SFO model \cite{Bugaev13}.
}
\label{Fig:sagunVIII}
\end{figure}

Also  we  found that the SFO$+\gamma_S$  fit leads to  a selective improvement and to  a certain degradation of the fit quality of various ratios for different collision energies. For instance, the  $\pi^-/\pi^+$ ratio is slightly  increased  for $\sqrt{s_{NN}} = $   6.3 and 7.6 GeV, but the situation drastically changes for $\sqrt{s_{NN}} = $ 12 GeV. The same tendency  is typical for  $\bar p/p$. On the contrary, for  $\bar \Xi^-/\Lambda$ ratio there  is a noticeably  worse data description within   SFO+$\gamma_s$ approach at $\sqrt{s_{NN}} = $  6.3, 7.6 GeV, but for $\sqrt{s_{NN}} = $ 12 GeV the fit quality is sizably better compared  to  all previous approaches. Thus, within the present model we  reveal a noticeable change in the trend  of   some ratios  at  $\sqrt{s_{NN}} = $  7.6-12 GeV, while at $\sqrt{s_{NN}} = $ 12 GeV we do not observe any sizable  improvement compared to the SFO model.

A special attention in our consideration  was paid to the Strangeness Horn, i.e. to the  $K^+/\pi^+$ ratio, because such a ratio  is  traditionally the most   problematic one for the HRGM to fit  it. As one can see  from Fig. \ref{Fig:sagunVI}, the remarkable  $K^+/\pi^+$  fit improvement  for $\sqrt{s_{NN}} = $ 2.7, 3.3, 4.3, 4.9, 6.3, 7.6, 12 GeV justifies the usage of the present  model. Quantitatively, we found that the $\chi^2/dof$ improvement for the SFO+$\gamma_s$ model  is  $\chi^2/dof$=1.5/14, i.e. even better than it was achieved  in \cite{Bugaev13} with $\chi^2/dof$=3.3/14 for the $\gamma_s$ fitting approach and  with  $\chi^2/dof$=6.3/14 for the  SFO model with  $\gamma_s=1$.

{
From Fig.  \ref{Fig:sagunVII} one can see that at two highest RHIC energies the description of ratios
within the single FO model is very good for all ratios except for $\bar \Lambda/\pi^-$ at $\sqrt{s_{NN}} = 130$ GeV and for $(\Omega+\bar \Omega)/\Xi^-$ at $\sqrt{s_{NN}} = 200$ GeV.  The main reason is that at these energies all chemical potentials except for the strange one are almost zero and hence the number of
particles and antiparticles is almost the same. As one can see from the upper panel of  Fig.  \ref{Fig:sagunVII},
the SFO$+\gamma_S$ fit significantly improves only  the
$\Omega/\pi^-$ and $\Xi^-/\pi^-$ ratios compared to the single FO model, i.e.  only two ratios of strange hadrons  responded to
the variation of two additional parameters. For $\sqrt{s_{NN}} = 200$ GeV  the SFO$+\gamma_S$ fit significantly improves only the $(\Omega+\bar \Omega)/\Xi^-$ ratio and worsens  the $\Lambda/\pi^-$ ratio and $\phi/p$ (less), i.e. only three ratios of strange hadrons  responded to such a sophisticated fit.
For $\sqrt{s_{NN}} = 62.4$ GeV  only two ratios out of five include kaons and, hence,  using $T_{\rm SFO}$ and $\gamma_s$ one can perfectly reproduce the strange particle ratios without affecting the non-strange ones.
Treating the $T_{\rm SFO}$ values found within the SFO model as an additional datum  for the SFO$+\gamma_S$ fit, we obtain  the same result.
Therefore, in contrast to low collision energies at high collision  energies only a few ratios  with strange particles  can be improved by simultaneous variation  of $T_{\rm SFO}$ and $\gamma_s$ and, hence, the SFO
cannot represent the majority of fitted ratios which are well reproduced even  within the single FO model with a single or with several hard-core radii.
The lower panel of Fig.  \ref{Fig:sagunIV} evidently  supports such a conclusion for the HRGM with the same value of hadronic hard-core radius.

Within   the SFO$+\gamma_S$ model the $\Lambda/ \pi^-$ and $\bar \Lambda/ \pi^-$  ratios  demonstrate some worsening  compared to less sophisticated models.
In Fig. \ref{Fig:sagunVIII} we show that   the SFO$+\gamma_S$ model  still  does not improve these ratios. The $\Lambda/ \pi^-$  fit quality, for instance, is  $\chi^2/dof$=10/8.  Hence, up to now  the best fit of  the $\Lambda/ \pi^-$ ratio  was obtained within  the SFO approach with $\gamma_s=1$. As it  was mentioned in \cite{KABAndronic:05,KABAndronic:09,KABugaev:Horn2013}
   a too  slow decrease of model results for $\Lambda/ \pi^-$ ratio   compared to the experimental data is typical for almost all statistical models. Evidently, the too steep rise in $\Lambda/ \pi^-$ behavior is a consequence of the $ \bar \Lambda$ anomaly \cite{KABAndronic:05,AGS_L3}. Similar results are reported in Refs. \cite{KABugaev:Becattini, KABugaev:Becattini13, KABugaev:Stachel}  as  the $\bar p$,  $\bar \Lambda $ and $\bar \Xi$ selective suppression. Since even an  introduction of the separate strangeness freeze-out with the strangeness enhancement  factor does not allow us  to better describe   these  ratios, we believe that  there is  a corresponding  physical reason which is responsible  for   it.
 One of them could be a necessity to introduce the different hard-core radius $R_\Lambda$ for the $\Lambda $ (anti)hyperons \cite{Sagun:2014}.
}

\section{Conclusions}

{
We present  a thorough investigation  of  the data measured at  AGS, SPS and RHIC energies within different versions of the multi-component hadron resonance gas model. The suggested  approach to  separately treat   the  freeze-outs of strange and non-strange  hadrons with the  simultaneous $\gamma_s$ fitting  gives  rise for the top-notch Strangeness Horn description with $\chi^2/dof$=1.5/14. The developed model   clearly  demonstrates  that the successful fit of  hadronic multiplicities  includes all the  advantages of these two approaches  discussed in  \cite{Bugaev13}.
As a result  for $\sqrt{s_{NN}} = $ 6.3, 7.6, 12, 130 GeV we found a significant data fit  quality improvement.

At the same time the lack of  available  data at
$\sqrt{s_{NN}}$ =2.7, 3.3, 3.8, 4.3, 9.2, 62.4 GeV forced us to redefine the fitting procedure at these collision energies in order
to avoid the mathematical inconsistency
which in combination with the  large experimental  error bars led to rather large uncertainties of  the fitting parameters.
The suggested redefinition of the fitting procedure by including the $T_{\rm SFO}$ temperatures obtained for these energies within the SFO model allowed us to avoid the mathematical problems and to get the safe answers on the values of the residual chemical non-equilibrium of strange particles. The developed sophisticated HRGM, i.e. the SFO+$\gamma_s$ model, allowed us
to  describe   the hadron multiplicity ratios with rather high quality $\chi^2/dof = 43.72/47 \simeq$ 0.93. This very  fact demonstrates  that the  suggested approach is a precise tool to elucidate the thermodynamics properties of hadron matter at two chemical freeze-outs. The fresh  illustrations  to this statement can be found in \cite{Bugaev13new}.

The achieved  total value of $\chi^2 = 43.72$ for the SFO+$\gamma_s$ model is almost 50\% lower than the $\chi^2$ value found for the single FO model and 30\% lower than the SFO model $\chi^2$  value.  The obtained  $\gamma_s$ values are  consistent with the conclusion  $\gamma_s \simeq 1$ (within the error bars). An evident  exception is the topmost  point of the Strangeness Horn (located at $\sqrt{s_{NN}} = $ 7.6 GeV), at which the mean value of the strangeness
enhancement factor is $\gamma_s \simeq 1.19 \pm0.15$. To reveal the physical reason for such a deviation we need
more experimental data with  an essentially  higher accuracy.

One of the main results obtained here is that the idea of separate chemical FO of strange hadrons provides a very high quality of  the data description. Further inclusion of the chemical non-equilibrium on top of the SFO is consistent with the result
$\gamma_s \simeq 1$ for all energies except for $\sqrt{s_{NN}} = $ 7.6 GeV.
Thus, the found residual chemical non-equilibrium of strange particles is weak  and, hence,   it  can be safely ignored for all energies except for $\sqrt{s_{NN}} = $ 7.6 GeV.  The  strange charge enhancement  of about 20\%   found at this collision energy allowed us to perfectly describe the topmost point of the Strangeness Horn, but at the expense of the worsening of  $\Lambda/ \pi^-$  and $\bar  \Lambda/ \pi^-$  ratios.

In addition, the description of ratios containing the   non-strange particles, especially such  as $\pi^-/\pi^+$ and $\bar p/p$,  gets better compared  to  previously reported  results \cite{KABugaev:Horn2013, Bugaev13}.  The remaining   problem of the ratios  involving  the  $\Lambda$   and  $\Xi$ (anti)hyperons can be resolved by
an inclusion of the different hard-core radius for $\Lambda$ (anti)particles \cite{Sagun:2014}, but such a treatment  is out of scope of the present work.

The performed analysis of the SFO+$\gamma_s$ model  hadronic pressures existing  at FO and at  SFO allowed us to elucidate an important  conclusion  that the single FO models with the same hard-core radius  \cite{KABAndronic:05}  or with different hard-core radii  \cite{KABugaev:Horn2013,Bugaev13,MultiComp:13} for all hadrons reproduce the SFO pressure for all collision energies below $\sqrt{s_{NN}} = 62.4$ GeV. The main reason for such a behavior is that the number of  ratios involving
strange hadrons is larger than the number of ratios with non-strange hadrons.

Also we report  an existence of strong jumps in the SFO pressure,  the SFO temperature  and  the corresponding effective number of degrees of freedom, when the center of mass collision energy changes from 4.3 to 4.9 GeV.
Based on the concept of non-smooth chemical freeze-out introduced recently in \cite{Bugaev13new}, we parameterized the dependencies
$T_{\rm FO}(\sqrt{s_{NN}})$ and  $T_{\rm SFO}(\sqrt{s_{NN}})$ which can be verified  in the future experiments
planned at FAIR (GSI) and NICA (JINR).
We hope that the high precision data measured in these experiments  will allow us to finally answer the question whether the residual non-equilibrium of strange charge is necessary to describe the topmost point of the Strangeness Horn or the concept of two separate chemical freeze-out for strange and non-strange hadrons can do this without introducing the $\gamma_s$ factor.
}

\vspace*{3mm}

\noindent
{\bf Acknowledgments.}  We would like to thank A. Andronic for  providing an access to well-structured experimental data.
The authors are thankful to  I. N. Mishustin, D. H. Rischke  and L. M. Satarov for valuable comments.
K.A.B., A.I.I.  and G.M.Z.  acknowledge  a  support
of  the Fundamental Research State Fund of Ukraine, Project No F58/04.
K.A.B.   acknowledges  also  a partial support provided by the Helmholtz
International Center for FAIR within the framework of the LOEWE
program launched by the State of Hesse.

\end{document}